\pdfoutput=1

\documentclass[10pt, conference]{IEEEtran}

\usepackage[noadjust]{cite}
\usepackage{multirow} 
\usepackage{algpseudocode}
\usepackage{algorithm}
\usepackage{rotating}
\usepackage{kantlipsum} 
\usepackage{commath}
\allowdisplaybreaks
\usepackage{mathtools}  
\usepackage{verbatim}
\usepackage{xr-hyper} 
\usepackage{enumitem}

\usepackage{epstopdf}
\usepackage{wrapfig}
\usepackage{latexsym}
\usepackage{amssymb}
\usepackage{amsthm}
\usepackage{amsfonts}
\usepackage{amsmath} 
\usepackage{graphicx}
\usepackage{latexsym}
\usepackage{booktabs}
\usepackage{support-caption} 
\usepackage{subcaption} 
\usepackage{xspace}

\usepackage[normalem]{ulem} 

\usepackage[T1]{fontenc}

\usepackage{listings}
\usepackage[dvipsnames]{xcolor}
\usepackage{multicol}
\usepackage{setspace}

\usepackage{booktabs}
\usepackage{graphicx}

\definecolor{background}{HTML}{EEEEEE}
\lstdefinelanguage{p4}{
  sensitive = true,
  backgroundcolor=\color{background},
  keywords=[2]{if, else, default, table},
  otherkeywords={\#define, \#undef, \#if, \#else, \#endif, \#ifdef, \#ifndef, \#elif, \#include},
  keywords=[3]{action, apply, bit, const, control, default, enum, error, extern, false, header,
    header_union, in, inout, int, package, parser, out, select, state, struct,
    transition, true, typedef, varbit, verify, metadata, header_type, fields, actions, key, default_action, counters
    },
  keywordstyle=\color{black},
  keywordstyle=[2]\color{black},
  keywordstyle=[3]\color{black},
  identifierstyle=\color{black},
  sensitive=false,
  comment=[l]{//},
  morecomment=[s]{/*}{*/},
  commentstyle=\color{ForestGreen}\ttfamily,
  stringstyle=\color{red}\ttfamily,
  morestring=[b]',
  morestring=[b]",
}

\ifCLASSINFOpdf
\else
\fi
\newcommand{\dq}[1]{``#1''}

\usepackage{comment}

\newcommand{\commentBy}[3]{\textcolor{#1}{\textbf{#2:} #3}}

\newif\ifcommentson

\newcommand{\ste}[1]{\ifcommentson \commentBy{blue}{SS}{#1} \fi}

\newcommand{\tulu}[1]{\ifcommentson \commentBy{red}{AT}{#1} \fi}

\newif\ifextended
\newif\ifshortver

\extendedtrue

\includecomment{extended-env}

\newcommand{\extended}[1]{\ifextended \ifshortver \textcolor{purple}{#1} \else #1 \fi  \fi}

\newcommand{\shortver}[1]{\ifshortver \ifextended \textcolor{blue}{#1} \else \textcolor{black}{#1} \fi \fi}

\newif\ifrevision

\begin{document}

\bstctlcite{IEEEexample:BSTcontrol}


\title{Micro SIDs: a solution for Efficient Representation of Segment IDs in SRv6 Networks}
\author{\IEEEauthorblockN{
Angelo~Tulumello\IEEEauthorrefmark{1},
Andrea~Mayer\IEEEauthorrefmark{1}\IEEEauthorrefmark{2},
Marco~Bonola\IEEEauthorrefmark{1}\IEEEauthorrefmark{2},
Paolo~Lungaroni\IEEEauthorrefmark{1}\IEEEauthorrefmark{2},
Carmine~Scarpitta\IEEEauthorrefmark{1}\IEEEauthorrefmark{2},\\
Stefano~Salsano\IEEEauthorrefmark{1}\IEEEauthorrefmark{2},
Ahmed~Abdelsalam\IEEEauthorrefmark{4},
Pablo~Camarillo\IEEEauthorrefmark{4},
Darren~Dukes\IEEEauthorrefmark{4},
Francoid~Clad\IEEEauthorrefmark{4},
Clarence~Filsfils\IEEEauthorrefmark{4}
}
\IEEEauthorblockA{
\IEEEauthorrefmark{1}University of Rome Tor Vergata,
\IEEEauthorrefmark{2}CNIT,
\IEEEauthorrefmark{4}Cisco Systems
}
}

\markboth{Submitted paper}%
{Author \MakeLowercase{\textit{et al.}}: Bare Demo of IEEEtran.cls for IEEE Journals}
%


\shortver{
\IEEEoverridecommandlockouts
\IEEEpubid{\makebox[\columnwidth]{978-3-903176-31-7\copyright2020 IFIP \hfill} \hspace{\columnsep}\makebox[\columnwidth]{ }}
}

\maketitle

\begin{abstract}
The Segment Routing (SR) architecture is based on loose source routing. A list of instructions, called segments can be added to the packet headers, to influence the forwarding and the processing of the packets in an SR enabled network. In SRv6 (Segment Routing over IPv6 data plane) the segments are represented with IPv6 addresses, which are 16 bytes long. There are some SRv6 service scenarios that may require to carry a large number of segments in the IPv6 packet headers. Reducing the size of these overheads is useful to minimize the impact on MTU (Maximum Transfer Unit) and to enable SRv6 on legacy hardware devices with limited processing capabilities that could suffer the long headers. In this paper we present the Micro SID solution for the efficient representation of segment identifiers. With this solution, the length of the segment list can be drastically reduced.
\end{abstract}


%

\IEEEpubidadjcol

\section{Introduction}

%
%
%
%
\IEEEPARstart{T}{he} SRv6 (Segment Routing over IPv6) Network Programming framework \cite{id-srv6-network-prog} extends the Segment Routing architecture \cite{filsfils2015segment,rfc8402}. According to \cite{id-srv6-network-prog}, a \textit{packet processing program} can be expressed with a sequence of instructions called \textit{segments}. Each instruction is encoded in a Segment ID (SID) which is 16-byte long (128 bits, the same size of an IPv6 address). SRv6 leverages the Segment Routing Header (SRH) \cite{rfc8754} to encode the packet processing program in the IPv6 packet headers as a \textit{Segment List}, together with optional metadata.

In SRv6 jargon, an operation to be executed at a node is called a \textit{behavior}. The packet processing instructions may express: i.) topological or traffic-engineering behaviours, such as \dq{go to this node via the Best-Effort Slice} or \dq{go to this node via the Low-Latency Slice}; ii.) fast-reroute behaviours, such as \dq{upon the sudden loss of a link, reroute the traffic via an optimum backup path}; iii.) VPN behaviours, such as \dq{egress the network via a specified Virtual Private Network (VPN) table of a specified Provider Edge (PE) router}. More in general, any application behaviour can be encoded in a network program, to be executed by a physical service appliance or by a softwarized component running in a virtual machine or in a container.

As discussed in \cite{id-shorter-srv6-sid-req}, some application scenarios for SRv6 may require long sequences of SIDs to be carried in the SRH packet header (e.g. up to 15 SIDs). In the current SRv6 model, this requires $N*16$ bytes to be carried in the SRH, where $N$ is the number of SIDs in the SID list. For this reason, an open research and technological problem is to find a solution to shorten the length of the SID representation in the packet headers. In this paper we present the \textit{Micro SID} solution \cite{id-srv6-usid}, its implementation in three different targets and a use case showing the interoperability among them. 

\ste{Stefano: per come ho scritto il testo sopra, forse possiamo solo aggiungere le references al posto di [xx,xx,...]. L'unica cosa da verificare e' la sezione IV. Se nella sezione IV confrontiamo uSID con altre soluzioni, allora dobbiamo introdurle qui}

The Micro SID solution introduces a straightforward extension to the SRv6 network programming model: each 16-byte SID can encode a micro-program rather than a single instruction. A micro-program is composed of micro-instructions, each represented with a \textit{Micro SID}, also called \textit{uSID}. 

\ste{Goals of the paper:
\\ 
\\- introduce the microsegment architeture
\\ 
\\- evaluate the savings
\\ 
\\- assess the interoperability of multiple implementations of SRv6 in a replicable testbed / experiment
\\ 
\\- show the easiness of the P4 implementation (we prove the low impact of the proposed extension by showing how easily the Micro SID mechanisms can be implemented in P4 by simply adding a few lines of code to a plain P4 SRv6 implementation)
\\ 
\\- evaluate the performance of software based implementation
}

\ste{paper content by section:}

In this paper we give a brief description of the SRv6 framework in Section \ref{sec:srv6} to explain the basic functionalities exploited in the Micro SID solution, presented in Section \ref{sec:arch}. In Section \ref{sec:saving} we analyze the saving in terms of header size compared to base SRv6 obtained with the Micro SID solution and with another proposed solution called SRm6 \cite{bonica-spring-sr-mapped-six}. We present the Micro SID implementation on Linux, VPP and P4 platforms in Section \ref{sec:design-implementation} and show the interoperability of the three implementations in Section \ref{sec:interop}. We evaluate the processing load performance of the Micro SID implementations in Section \ref{sec:assessment} and discuss related works in Section \ref{sec:related}.

\section{SRv6 Network Programming framework}
\label{sec:srv6}

In this section, we shortly recall the main features of SRv6 Network Programming framework, as needed to understand the rest of the paper. For further details, we refer the reader to the specification of the framework in \cite{id-srv6-network-prog} and to the tutorial on SRv6 that is available in \cite{ventre2019segment}.

An SRv6 SID can be partitioned in three parts and expressed as LOC:FUNCT:ARG (Locator, Function, Argument). The Locator part can be routable and used to forward a packet to a specific node, where a behavior, identified by the Function part needs to be executed. In most cases, the Argument part is not used, hence a SID can be simply decomposed in two parts LOC:FUNCT (Locator and Function). To provide an example (taken from \cite{id-srv6-network-prog}) an operator can use a /48 IPv6 network prefix for its SRv6 transport domain which include all SRv6 capable transport nodes.We refer to this prefix as \textit{Locator Block}. Each SRv6 capable node can be assigned a different /64 IPv6 network sub-prefix inside the Locator Block, therefore up to $2^{16}=65356$ SRv6 nodes can be supported in this specific configuration. Inside each SRv6 node, $2^{64}$ different SIDs can be supported. As an example (see Fig.~\ref{fig:un_example}, the /48 Locator Block prefix can be \texttt{fc00:1234:abcd::/48}, a specific node prefix can be \texttt{fc00:1234:abcd:N::/64}, and the SID of a behavior to be executed in the node can be \texttt{fc00:1234:abcd:0100::S}. In this case, the locator part (LOC) is represented by the leftmost 64 bits, composed by the Locator Block and by a node part N. In the example, the locator for node $R_N$ is \texttt{fc00:1234:abcd:0N00}. The FUNCT part is represented by the rightmost 64 bits (no ARGS is considered). In the example, \texttt{0001} or \texttt{F001} are used (preceded by 12 more leading zeros in hexadecimal notation). 

The regular routing protocols can be used to distribute the reachability information for the Locators associated to the SRv6 network nodes. In this way, a single routing prefix can be used to reach a given node and forward the packets towards all behaviors that can be executed by that node. To ease the interoperability, a set of \dq{well-known} behaviors is defined in \cite{id-srv6-network-prog} (but other documents can define additional behaviors). The most important SRv6 behaviors defined in \cite{id-srv6-network-prog} are briefly described hereafter.


The simplest SRv6 behavior is the \textit{End} behavior, which is used to enforce a topological waypoint in the path of a packet towards its final destination. In the example shown in Fig.~\ref{fig:un_example}, a packet coming from Site A enters the SR domain in node $R_1$, where it is encapsulated in an IPv6 outer packet. Starting from node $R_1$, the packet needs to cross $R_8$, then $R_7$, then it needs to reach $R_2$ where it will be decapsulated and sent to Site B. Each node $R_N$ advertises a /64 prefix, in the example \texttt{fc00:1234:abcd:0N00::/64}. Considering node $R_8$, the \texttt{fc00:1234:abcd:0800::0001} SID is mapped into the \textit{End} behavior in node $R_8$ reached with the \texttt{fc00:1234:abcd:0800::/64} prefix. The End behavior simply corresponds to \dq{consuming} one SID in the SID list, therefore node $R_8$ will read the next SID in the SID list and will update the IPv6 destination address with the next SID. The End.X behavior is meant to cross-connect the packet towards a specific next hop. The End.T behavior is used to use a specific routing table for the the IPv6 route lookup (as needed for example to implement VPNs with per-customer routing tables). The End.DX6 behavior is used to decapsulate a packet, extracting it from the outer IPv6 packet, and to cross connect it to a specific IPv6 next hop. The End.DT6 behavior is used to decapsulate a packet and then to use a specific routing table for the IPv6 route lookup of the inner packet.

\subsection{SRv6 Control Plane aspects}

An operator is free to associate a SID (logically split into LOC:FUNCT or LOC:FUNCT:ARGS) to a given behavior in a given node. The specific values for the SIDs and in particular for the FUNCT part can be provisioned and managed by an SDN controller, and/or they can be advertised by routing protocols (OSPF, ISIS, BGP) with SRv6 specific extensions. We observe that by using an SDN based approach, the use of SRv6 specific routing protocol extensions is optional. An SRv6 network can be operated by only distributing node reachability information (regular IPv6 prefixes) in routing protocols, assuming that an SDN controller manages the association of node SRv6 behaviors to SID values. 

\section{Micro SIDs}
\label{sec:arch}

\begin{figure}[t]
	\centering
	\includegraphics[width=\columnwidth]{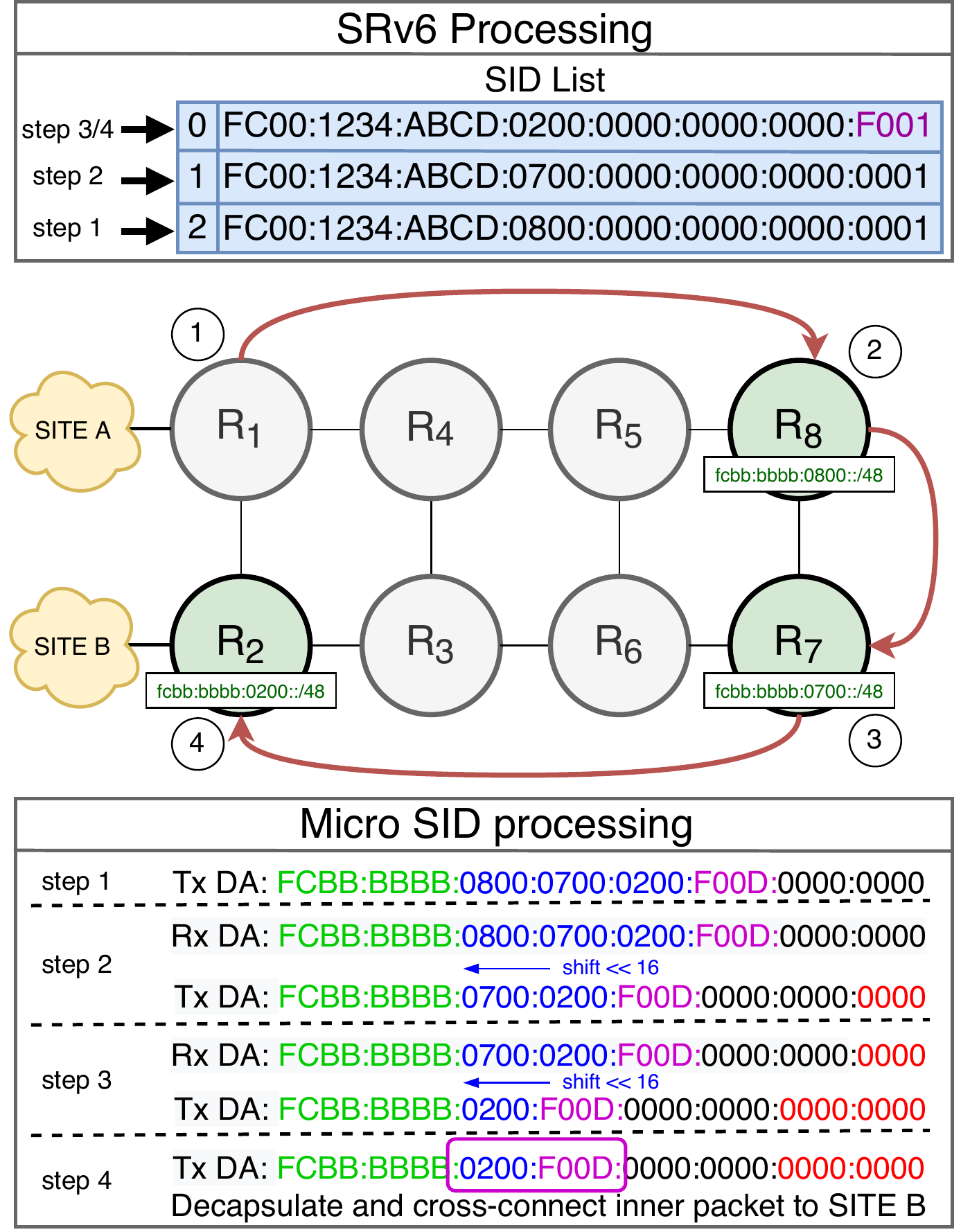}
	\caption{Plain SID and Micro SID example}
	\label{fig:un_example}
\end{figure}

The fundamental idea of the Micro SID solution \cite{id-srv6-usid} is that each 16-byte instruction (SID) of an SRv6 packet can carry a micro-program, composed of micro-instructions represented with identifiers called Micro SIDs. This approach results in a large saving of the packet overhead when multiple segments (instructions) needs to be transported in an SRv6 packet. \extended{Considering the most common configuration, Micro SIDs are represented with 2 bytes, and up to 6 Micro SIDs can be carried in a regular 16 bytes SID. For example, an SR path of 5 hops would require 16*5=80 bytes to be represented as a regular \dq{SR path}, while it would require 16 bytes to be represented as a sequence of 5 Micro SIDs encoded in a single regular SID.
A micro-instruction is represented with a \textit{Micro SID}, also called \textit{uSID}.} In this work we will consider that uSIDs are represented with 2 bytes, but other choices are possible (e.g. using 3 or 4 bytes).

\begin{table}
\centering
\begin{tabular}{|c|c|}
\hline
\textbf{Plain SRv6} & \textbf{Micro SID}                           \\ \hline
End                & uN                                         \\ \hline
End.X                & uA                                        \\ \hline
End.DT4/End.DT6/End.DT2                 & uDT                    \\ \hline
End.DX4/End.DX6/End.DX2                 & uDX                     \\ \hline
\end{tabular}%
\caption{Plain SRv6 behaviors and Micro SID behaviors}
\label{tab:behavior-functions}
\end{table}

As described in \cite{id-srv6-usid}, the Micro SID solution proposes to extend SRv6 Network Programming with new behaviors, called uN, uA, uDT, uDX, as described in Table \ref{tab:behavior-functions}.

To introduce the reader to the basic Micro SID processing, we describe a simple use case example, based on the same reference topology of the SRv6 example, depicted in Figure \ref{fig:un_example}. In this case a /32 prefix is chosen as Locator Block for the Micro SIDs (referred to as \textit{uSID block}). All routers in the topology are assigned a /48 prefix from this Micro SID block: \texttt{fcbb:bbbb::/32}. The ingress router R1 applies the Micro SID policy by encoding the address \texttt{fcbb:bbbb:0800:0700:0200:f00d::} into the outer IPv6 header. This results into a source routing policy that routes the packet through the path $R_8\rightarrow R_7\rightarrow R_2$, respectively identified by the Micro SIDs \texttt{0x0800}, \texttt{0x0700} and \texttt{0x0200} and then executes a \textit{decap} operation. Thus, $R_1$ sends the packet to $R_8$. The packet will cross $R_4$ and $R_5$ that in this case enforce \dq{base} IPv6 forwarding. As soon as $R_8$ receives the packet, it \dq{consumes} its Micro SID identifier in the destination address: (i) the \texttt{0x0800} Micro SID is popped from the destination address; (ii) the remaining Micro SID list is shifted left by 16 bits; (iii) the End of Container identifier (\texttt{0x0000}) is inserted in the last 16 bits. The resulting IPv6 destination address is \texttt{fcbb:bbbb:0700:0200:f00d::}.  Upon completion of the procedures above, the packet is transmitted to $R_7$ which performs an analogous set of procedures that ends with the transmission of a packet containing the Micro SID list \texttt{fcbb:bbbb:0200:f00d::} to $R_2$ via $R_6 \rightarrow R_3$. Since $R_2$ is the last SRv6 router in the path, the destination address of the packet matches the FIB entry with destination \text{fcbb:bbbb:0200:f00d::/64}. This rule includes the terminator Micro SID \texttt{f00d} which triggers the final End.DT6 behavior: the packet is decapsulated and handled by a specific IPv6 routing table. 
 
The Micro SID solution fully leverages the SRv6 network programming solution. In particular, the data plane with the SRH dataplane encapsulation is leveraged without any change; any SID in the SID list can carry micro segments. As for the Control Plane, the SRv6 Control Plane is leveraged without any change. The mechanisms for the compression of SID identifiers are described in \cite{id-srh-comp-sl-enc}.

The Micro SID solution enables ultra-scale deployments (e.g. as needed for multi-domain 5G scenarios) and reduces the overhead at the minimum reducing the potential issues with MTU. It is fully compatible with SRv6 architecture, so it can run in mixed scenarios where only a subset of nodes support the Micro SIDs.

\section{Evaluation of compression savingss}
\label{sec:saving}
This section provides a detailed analysis of the efficiency of the Micro SID compression in a realistic SRv6 deployment scenario. In particular, it considers the encapsulation size of a compressed segment lists versus an uncompressed segment list. The efficiency of the Micro SID solution is also compared with another proposed SRH compression solutions called SRm6 (Segment Routing Mapped to IPv6) \cite{bonica-spring-sr-mapped-six}.

We show that a mapping solution (like SRm6) does not provide better compression than what can be achieved with the SRv6 mechanism. As such, analysis of the SRm6 proposal documented in \cite{bonica-spring-sr-mapped-six} is provided for comparison.


The SRm6 solution \cite{bonica-spring-sr-mapped-six} defines a new routing header called \textit{Compact Routing header} to be used to carry the list of segments instead of the SRH. More specifically, \cite{bonica-spring-sr-mapped-six} defines two versions of CRH: CRH-16 and CRH-32, that respectively support Segment Identifiers (SIDs) of 16 bits (2 bytes) and 32 bits (4 bytes). The SRm6 SIDs needs to be \textit{mapped} into IPv6 addresses, locally on each node of an SRv6 network. A \dq{Per Path Service Instruction} can be encoded in a new Option to be included in a \textit{Destination Option} header of the IPv6 packet.

Note that the uSID solution is fully compatible at the data plane level with the SRv6 framework, as the packet forwarding is based on IPv6 Destination Addresses and on the SRH. The SRm6 requires a data plane based on a new Routing Header and on a new Option in the Destination Option header. 

\subsection{Reference topology and scenario}
\begin{figure}[t]
	\centering
	\includegraphics[width=\columnwidth]{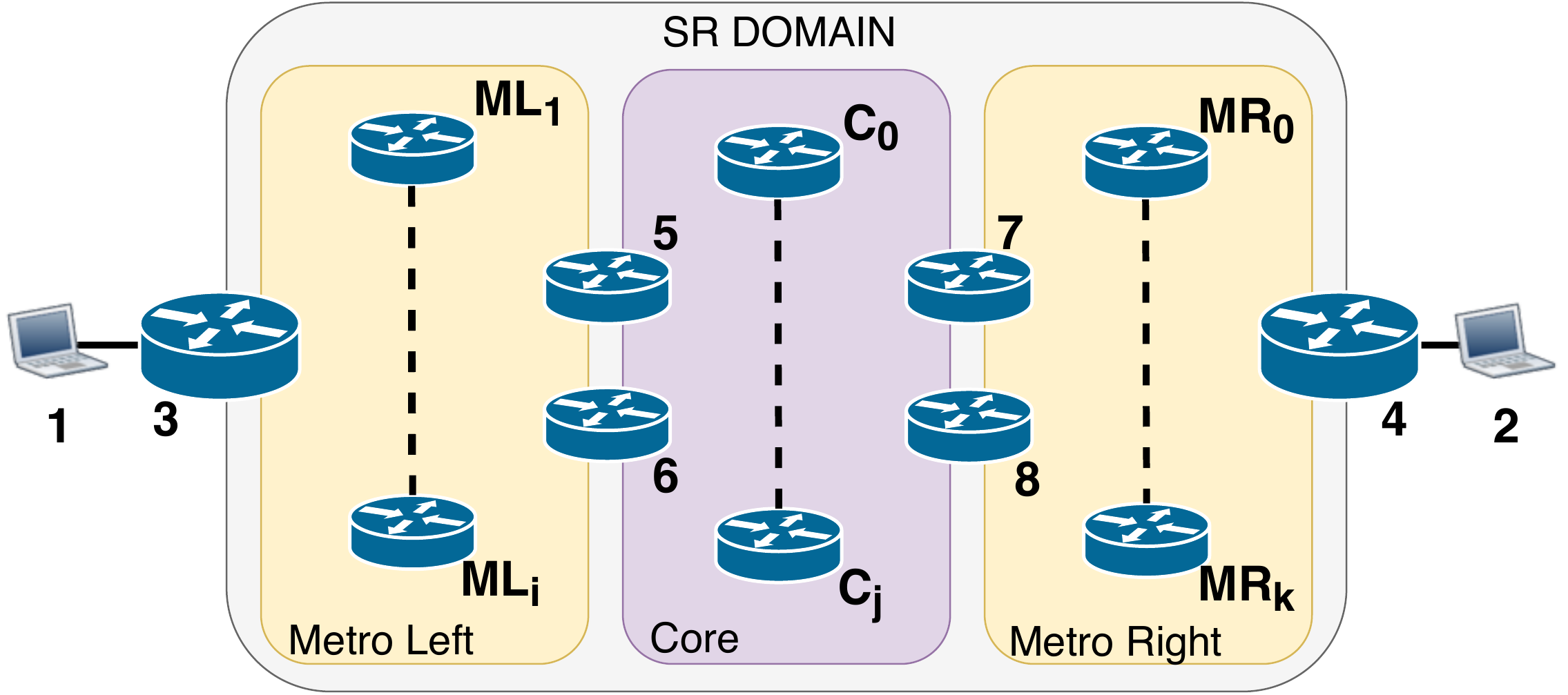}
	\caption{Reference topology for Compression Analysis}
	\label{fig:compression-topo}
\end{figure}

Let us consider a service provider offering a VPN service with underlay optimization. The reference topology is depicted in  Fig.~\ref{fig:compression-topo}. Hosts 1 and 2 are located in two different sites of a VPN customer. When host 1 sends a packet to host 2, the SR domain ingress router 3 steers it to the egress edge router 4 via an SR Policy that enforces a path through a number of underlay waypoints in Metro L ($\mathrm{ML_1}$..$\mathrm{ML_i}$), Core ($\mathrm{C_1}$..$\mathrm{C_j}$), and Metro R ($\mathrm{MR_1}$..$\mathrm{MR_k}$).  The SR Policy ends with a SID that instructs the egress edge router 4 to decapsulate the packet and forward it towards host 2.
   
\subsection{Compression Analysis}
In the following, we analyze and compare the header lengths of the uSID and SRm6 with respect to the basic SRv6 header. In particular, we evaluate the \dq{Encapsulation size Saving} i.e. the fraction of Encapsulation overhead that is saved using a compression solution with respect to the original (uncompressed) Encapsulation overhead introduced by the SRv6 solution based on full IPv6 SIDs and SRH.

According to \cite{filsfils-spring-analysis-fmwk-ext-srv6-encap}, we define the Encapsulation size metric $E(SL)$ as the number of bytes required to encapsulate a packet traversing an SRv6 domain. It includes all the bytes of the \dq{outer} IPv6 packet, from the beginning of the outer IPv6 packet (at layer 3) up to the beginning of the encapsulated packet. We note that the encapsulation size $E(SL)$ is a function of the Segment List Size $SL$, as each Segment in the SID List needs to be represented in the outer IPv6 packet.

The value of the the Encapsulation size metric is calculated for reduced SRv6 encapsulation as $E(SL) = 40$ bytes (IPv6 Header) if $(|SL| = 1)$ or $(E(SL) = 40 + 8 + (|SL| - 1) * 16)$ otherwise. Where 40 is the IPv6 header, 8 is the fixed part of SRH and 16 is the size of IPv6 address.

The SRv6 basic encapsulation is evaluated considering the reduced encapsulation policy ($H.Encap.Red$), defined in \cite{id-srv6-network-prog} section 5.1. The $H.Encap.Red$ policy encapsulates an IPv6 packet into an outer IPv6 packet with the SRH header. The first SID of the segment list is placed in the IPv6 Destination Address of the outer IPv6 packet and is not replicated in the SRH. If the SID list consists of only one SID, the entire SRH header may be omitted, resulting in a plain an IPv6 in IPv6 packet without the SRH extension header.

According to \cite{dukes-srv6-overhead-analysis}, we define the Encapsulation size Saving $ES$ metric considering the Encapsulation size of the compressed solutions $E_{c}(SL)$ and the Encapsulation size of plain SRv6 without any compression encoding $E_{p}(SL)$, as follows: \(ES(SL) = 1 - E_{c}(SL) / E_{p}(SL) \).

For the analysis of the Micro SID solution, the Locator Block identifies the SRv6 domain, while the the Node\&Function Block represents the node identifier along with the function to be applied. The Argument block contains the metadata needed to carry out the behavior processing.

A 16-byte SRv6 instruction that contains a micro-program is called a uSID \textit{container} instruction and has the structure shown in Fig.~\ref{fig:un_example}. We measure the capacity $C_{uSID}$ of a uSID container as follows:
  
\begin{equation*}
C_{uSID} = \left \lfloor \frac{(128 - B)}{NF} \right \rfloor
\end{equation*}
where $B$ and $NF$ are the lengths of the Locator and the Node\&Function blocks, respectively.

Given a sequence S of uncompressed SIDs the length of the corresponding uSID sequence is evaluated as follows:

\begin{equation*}
L_{uSID}(S) = \left \lceil \frac{\left | S \right |}{C_{uSID}} \right \rceil
\end{equation*}

For SRm6, SIDs of fixed size are used, of 16 or 32 bits which are carried in a new Routing Header called Compact Routing Header (CRH) \cite{bonica-6man-comp-rtg-hdr-22}. The CRH is made of a fixed set of fields (i.e NextHdr, HdrLen, RoutingType, SegmentLeft, SID[0], SID[1]) for a total of 8 bytes and a variable length list of SIDs. The CRH must end on a 64-bit boundary otherwise it must be padded with zeros.

SRm6 expects headers with 16-bit or 32-bits defined as CRH-16 and CRH-32, respectively. In CRH-16 the length of headers is \(E_{CRH16}(SL) = 40 + 8\) if \(|SL| = 1\), otherwise \(E_{CRH16}(SL) = 40 + \lceil (4 + |SL| * 2) / 8 \rceil * 8 + 8 \).

In CRH-32 the length of headers is calculated as \(E_{CRH32}(SL) = 40 + 8 \) if \(|SL| = 1\), otherwise \(E_{CRH32}(SL) = 40 + \lceil(4 + |SL| * 4) / 8\rceil * 8 + 8\).
\ste{TODO: qui non abbiamo spiegato bene queste formule sopra... da dove vengono i +8 alla fine? da dove vengono i 4 che vengono sommati in modo fisso?}
   

In our comparison, the uSID solution is considered with 16-bit uSID length (the uSID Block size is 32 bit). The SRm6 is considered with both CRH-16 and CRH-32 routing headers.

\begin{figure}[t]
	\centering
	\includegraphics[width=\columnwidth]{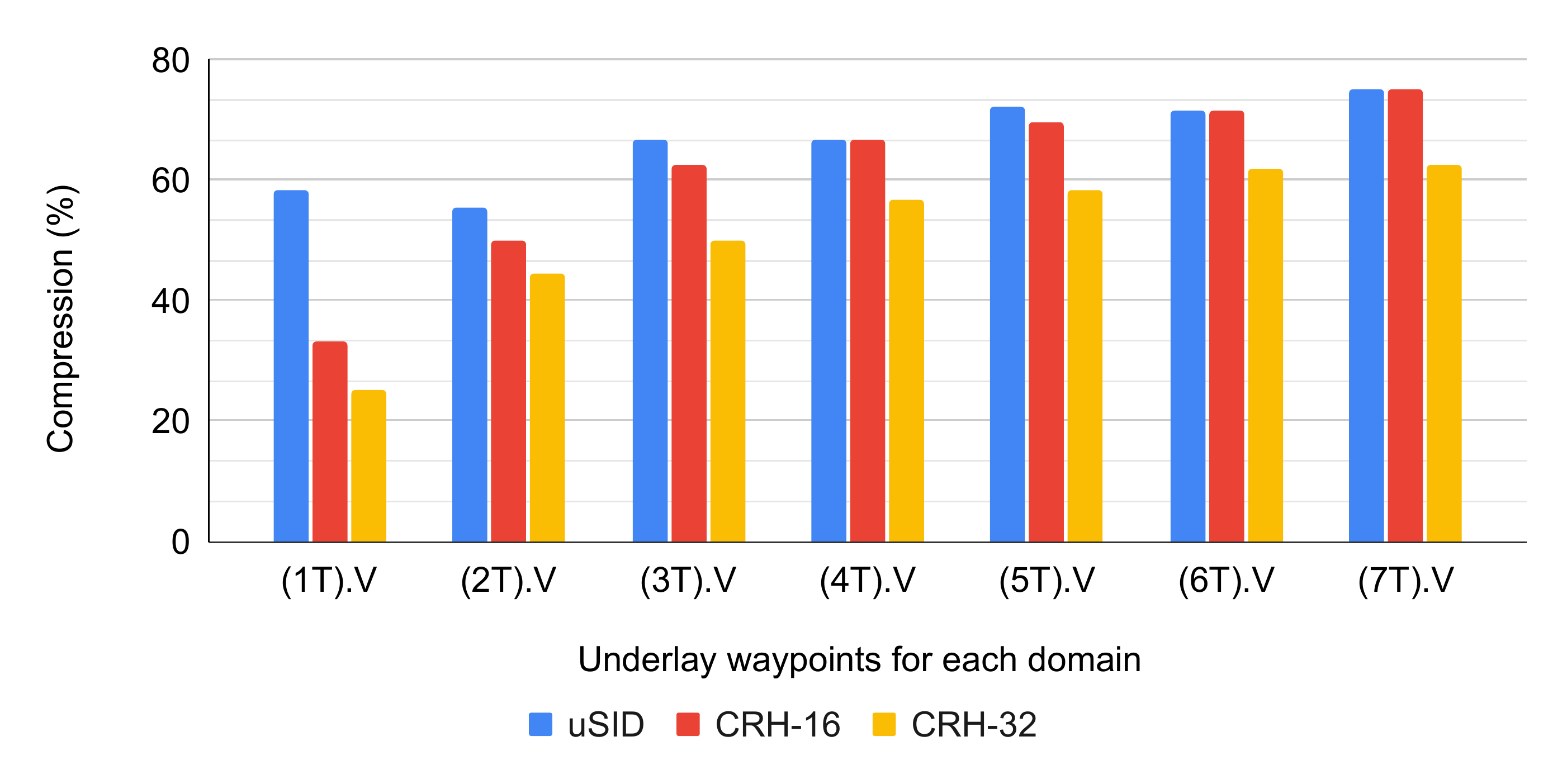}
	\caption{Encapsulation size Saving for uSID and SRm6}
	\label{fig:compression-analysis}
\end{figure}

Figure \ref{fig:compression-analysis} plots the Encapsulation size Saving for the three solutions, considering the reference scenario in a range from one to seven underlay waypoints for each domain (Metro L, Core, Metro R). The Micro SID compression is significant (58\%) also for a SID list of just 4 nodes (first group of bars in Figure \ref{fig:compression-analysis}), thanks to the Reduced Encapsulation in IPv6 that encodes the uSID list only in the IPv6 destination address, without adding the SRH with the SID list in the IPv6 header. Then, as the number of SIDs increases the uSID and the CRH-16 solutions align to a compression percentage around 70\%.

\section{Design and implementation}
\label{sec:design-implementation}

\subsection{Implementation of uSIDs using P4 Language}
A proof of concept implementation of uSID  primitives has been realized in P4, by extending a publicly available implementation of the SRv6 framework \cite{srv6-cascone}. To this end, we have developed the following extensions:

\begin{itemize}
    \item added a new action named \textbf{usid\_un}, responsible for (i) extracting the uSID of the next end router and (ii) updating the IPv6 destination address accordingly \shortver{(see the Listings reported in the extended version)}\extended{(see listing \ref{listing:uSID_un})}
    \item added a new Longest Prefix Match (LPM) table named \textbf{my\_usid\_table} responsible for the uN behavior \shortver{(see listing reported in the extended version)}\extended{(see listing \ref{listing:my_usid_table})}
    \item modified the overall application logic (i.e.: the P4 apply block) to invoke the new processing primitives
\end{itemize}
The full P4 implementation is available in our public repository \cite{usid-testbed-repo}. \shortver{Some P4 code listings and further details are available in the extended version \cite{usid-arxiv}.}


\begin{extended-env}

\begin{minipage}{\columnwidth}
\begin{lstlisting}[language={c},label=listing:uSID_un, xleftmargin=0.15cm, xrightmargin=0.35cm, caption={uSID\_un P4 code},captionpos=b]
#define USID_BLOCK_MASK 0xffffffff << 96
action usid_un() { 
    hdr.ipv6.dst_addr = (hdr.ipv6.dst_addr & USID_BLOCK_MASK) | ((hdr.ipv6.dst_addr << 16) & ~((bit<128>)USID_BLOCK_MASK));
}
\end{lstlisting}
\end{minipage}

\begin{minipage}{\columnwidth}
\begin{lstlisting}[label=listing:my_usid_table, xleftmargin=0.15cm, xrightmargin=0.35cm,
caption={my\_usid\_table P4 code},captionpos=b]
direct_counter(CounterType.packets_and_bytes) my_usid_table_counter;
table my_usid_table {
    key = {
        hdr.ipv6.dst_addr: lpm;
    }
    actions = {
        usid_un;
        end_action;
        NoAction;
    }
    default_action = NoAction;
    counters = my_usid_table_counter;
}
\end{lstlisting}
\end{minipage}

\end{extended-env}

To support the uN behavior, the implemented P4 pipeline requires two kinds of match/action entries. The first one matches on a /48 IPv6 prefix (e.g. \texttt{fcbb:bbbb:0100::/48}) and invokes the \texttt{usid\_un} action performing the shift-and-lookup primitive. The second one matches on a /64 prefix (e.g. \texttt{fcbb:bbbb:0100::/64} and triggers the SRv6 \texttt{End} behavior, i.e. decrement the SRH \texttt{segment\_left} field and copy the next SID from the SRH to the IPv6 destination address.

\subsection{Linux kernel uSID implementation}
\extended{Since release 4.14, the Linux kernel basically offers a subset of the behaviors defined in [draft-srv6] implementing SRv6 End, End.X, etc. However, the Linux kernellacks of the SRv6 uSID capabilities and thus it does not support any micro segment extension for SRv6 proposed in \cite{id-srv6-usid}.}

In order to add the support for uSID in the Linux kernel(which from release 4.14 already supports a subset of the SRv6 behaviours), we designed and implemented a patch that extends and enhances the existent SRv6 subsystem. The proposeed uSID implementation comes up with the support for the uN and uA behaviors which are, respectively, a variant of the Endpoint (End) and of the Endpoint with Cross Connect (End.X). Moreover, we have also extended the userspace \texttt{iproute2} suite \cite{iproute2} to support the new uSID behaviors. In particular, using the \texttt{ip} command we are able to instantiate and destroy instances of uN and uA behaviors.

\extended{
During the design and the development of the uSID patch set, we faced some architectural limitations of the Linux SRv6 implementation which are not due to our changes but they are intrinsic of the SRv6 subsystem itself.}

\extended{
Such limitations mainly concern the way in which behavior instances are created, configured and destroyed. SRv6 behavior instances are managed through the help of an userspace application which is in charge of sending to the kernel(through netlink messages) the commands for carrying out the requested operations.}

\extended{
In our case, the reference userspace application is represented by the ip command made available by the iproute2 suite. The ip command allows to create a new instance of a SRv6 behavior by specifying the type (i.e.: End, End.X, etc) and a list of per-behavior attributes (i.e: table id, nexthop, etc) used during setup phase. Similarly, the ip command also allows to destroy a specific SRv6 behavior instance which should release all the resources that it has requested (if any) during the creation/setup phase.
}

All the SRv6 behaviors implemented in the Linux kernel share the same basic creation/setup function whose purpose consists of allocating the memory for the new behavior instance and parsing the supplied attributes. First, the basic creation/setup function do not allow to specify a custom callback on a per-behavior basis that could be used for carrying out any sort of interaction with the rest of the kernelor allocating some additional memory. Second, such basic approach does not support optional attributes supplied by the userspace (which are required by the new uSID behaviors).

\extended{
Moreover, the SRv6 Linux implementation handles any supplied attribute as required. This means that the userspace must always provide all the attributes required by a specific behavior type otherwise the instantiation of such behavior fails.}

\extended{
Consequently, none of the different SRv6 behaviors implemented so far in the Linux kernel can support 1) custom creation/destroy callbacks and/or 2) it can manage optional attributes supplied by the userspace.
}

In order to implement the uN and uA behaviors we had to overcome the two above mentioned limitations. To this end we have: (1) extended the SRv6 implementation introducing two per-behavior callbacks which are called (if provided) when a new behavior instance is created and when it is going to be destroyed; (2) patched the SRv6 Linux kernelto support optional attributes for SRv6 behaviors without breaking any backward compatibility.

\extended{
Indeed, none of the different SRv6 behaviors implemented so far in the Linux kernelcan support custom creation/destroy callbacks (required to interact with the rest of the kernelor allocating some additional memory). Moreover, the current Linux SRv6 framework can not manage optional attributes supplied by the userspace when instantiating a specific behaviours. 
}

\extended{
Indeed both the aforementioned behaviors require of custom creation/destroy callbacks as well as the support for optional attributes.
}

\extended{
To overcome the limitation (1), we extended the SRv6 implementation introducing two per-behavior callbacks which are called (if provided) when a new behavior instance is created and when it is going to be destroyed.
}

\extended{
We got rid of the limitation (2) by patching the SRv6 Linux kernelto support optional attributes for SRv6 behaviors without breaking any backward compatibility. Before this patch if an attribute was not needed by a behavior, it required to be set by the userspace application to a conventional skip-value (otherwise the creation process fails if the attribute is not supplied).
}

\extended{
The kernel side, that processes the creation request of a behavior, reads the supplied attribute values and it checks if it had been set to the conventional skip-value or not. Hence, each optional attribute is supposed to have a conventional skip-value which is known a priori and shared between userspace applications and kernel. Hence, messy code and complicated tricks can arise from this approach. 
}

\extended{
The patch for optional attributes explicitly differentiates the required mandatory attributes from the optional ones. Now, each SRv6 behavior can declare a set of required attributes and a set of optional ones. The creation of a behavior instance fails in case a required attribute is missing, while it goes on without generating any issue if an optional attribute is not supplied by the userspace application.
}

\extended{
The uN and the uA patches share most of the code. The only difference between the two relies in the way in which the packets are forwarded when no more uSID are available in the IPv6 destination address. In this case, if the packet contains an SRv6 header,  the uN applies the SRv6 End behavior while the uA applies the End.X behavior. For this reason we do not treat the uN and the uA implementations separately. Therefore, in the following we will refer only to the uN implementation and we will describe the  most significant differences with respect to the uA behavior, when needed.
}

\extended{
First of all we extended the \texttt{enum} structure that defines all the SRv6 behavior types currently available in the Linux kerneladding the \texttt{SEG6\_LOCAL\_ACTION\_UN} and the \texttt{SEG6\_LOCAL\_ACTION\_UA} in the \texttt{include/uapi/linux/seg6\_local.h} header file.
We also defined the new SRv6 behavior attribute type \texttt{SEG6\_LOCAL\_USEG}
and the related data structure named \texttt{usid\_layout} which is used for passing the uSID block and the uSID lengths from the userspace to the kerneland vice versa.
}

\extended{As for the all existing SRv6 behaviors, the uN and uA implementations take place in the \texttt{net/ipv6/seg6\_local.c} file. In the Linux kernel, each SRv6 behavior is realized as an extension of a lightweight tunnel  (LWT) \cite{SRV6implem} through the \texttt{seg6\_local\_lwt} structure. We extended this structure to store both the uSID block and the uSID lengths provided by the userspace. Subsequently, we added the new uN and the uA behavior types to the descriptor of behaviors, \texttt{seg6\_action\_table},  specifying: i) the behavior type, ii) the optional uSID block and uSID lengths, iii) the processing function to be called for every incoming packet, iv) the setup callback to use when a new instance of a uN/uA behavior must be created.}

The source code of the patches to the kernel and to the iproute2 suite are available in ou project repository \cite{usid-linux-impl-repo}. \shortver{A more detailed explanation of the proposed uSID implementation can be found in the extended version of this paper \cite{usid-arxiv}.}

\extended{
When an IPv6 packet is received on a interface, it goes through the Linux networking stack and it reaches the input routing routine. At this point, the system performs the route lookup operation trying to determine the fate of the packet. As a result, the packet could be 1) discarded or 2) locally delivered or 3) forwarded to the next hop or 4) processed by an SRv6 behavior. 
In case of 4), the processing function depends on the SRv6 behavior type: for the uN the processing we defined the \texttt{input\_action\_un} function while for the uA we defined the \texttt{input\_action\_ua} function.
}

\extended{
Both the uN and the uA process an incoming IPv6 packet in the same way if the next uSID is not equal to the zeroed container and thus they share the same processing logic. In this case, the processing function consumes the current uSID replacing it with the next uSID. This operation is performed by the \texttt{un\_next\_usid} function which takes care of advancing the next uSID in the IPv6 destination address.
After that, the \texttt{seg6\_lookup\_nexthop} function carries out the route lookup using the updated uSID. Based on the result of the operation, the packet could be discarded or locally delivered or forwarded to the next hop.
On the contrary, when the next uSID is equal to the zeroed container (the current uSID is the last one in the IPv6 destination address), the uN performs the SRv6 End behavior while the uA performs the SRv6 End.X.
}

\extended{
The uN and the uA behaviors share the same setup function, the \texttt{input\_action\_ux\_build}, which is called during the creation of a new behavior instance. At this time, this function checks if the user supplied custom uSID block and uSID lengths. If the users decided to stay with the default values (none of these two attributes are given) the kernelreads the default uSID block and uSID lengths which can be set from the userspace through the sysctl command [add sysctl  paths here]. Otherwise, the kernelretrieves the missing attributes from the sysctl as for the default case. In both cases, the \texttt{input\_action\_ux\_build} always checks for the validity of the uSID block and the uSID lengths. Such values must be expressed in bits and they must be divisible by 8. Moreover, the function also checks for the semantic of the uSID block and uSID lengths by considering the fact that each IPv6 address must contain the uSID block, one uSID at least and the zeroed container.
}

\subsection{uSID VPP implementation}

Virtual Packet processor (VPP) is an open source virtual router \cite{vpp}. It implements a high-performance forwarder that can run on commodity CPUs. VPP often runs on top of the Data Plane Development Kit (DPDK) \cite{dpdk} to achieve high speed I/O operations. DPDK maps directly the network interface card (NIC) into user-space bypassing the underlying Operating System kernel.

\extended{
\begin{figure}[htbp]
	\centering
	\includegraphics[width=\columnwidth]{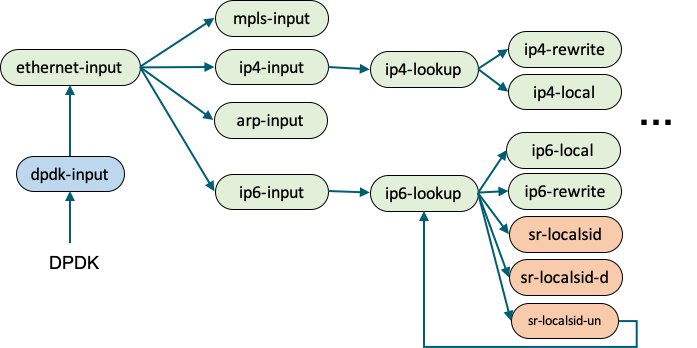}
	\caption{VPP Packet Processing Architecture}
	\label{fig:vpp-pkt-process}
\end{figure}
}

The packet processing architecture of VPP consists of graph nodes that are composed together. Each graph node performs one function of the processing stack such as IPv6 packets input (\emph{ip6-input}), or IPv6 FIB look-up (\emph{ip6-lookup}). The composition of the several graph nodes of VPP is decided at runtime. \extended{Fig.~\ref{fig:vpp-pkt-process} shows an example of a VPP packet graph.} VPP supports most of the behaviors defined in~\cite{id-srv6-network-prog}. \extended{The behaviors are implemented in the \emph{sr-localsid} and \emph{sr-localsid-d} VPP graph nodes.} 

We added a new VPP graph node (\emph{sr-localsid-un}) to support the SRv6 uSID uN behavior. The new VPP graph node implements the shift-and-lookup functionality. When a new is uN behavior is created using VPP CLI/API, two separate FIB entries are created. The first FIB entry (e.g., \texttt{FC00:0000:0100::/48}) triggers a shift-and-lookup of the IPv6 destination address, while the second FIB entry (e.g., \texttt{FC00:0000:0100::/64}) triggers the SRH processing (implemented in the \emph{sr-localsid} VPP graph node) by copying the next 128b SID from the SRH to the IPv6 destination address. 
 
A received SRv6 packet may match either of the two FIB entries. Depending on which FIB entry the packet hits, it gets processed by a different VPP graph node. In this way we maintain the VPP performance and avoid instruction cache misses as all the packets that arrive to the VPP graph node must execute the same instruction, being either shift-and-lookup or SRH processing.
\extended{As in any SRv6 functionality, the \emph{sr-localsid-un}) graph node also maintain counters to track traffic destined to such SID. Once the shifting is completed the packet is redirected to the (\emph{ip6-input}) VPP graph node to determine the egress interface for the packet based on the updated IPv6 destination address.} 

\section{Interoperability and Testbed}
\label{sec:interop}


\extended{
\begin{figure*}[t]
	\centering
	\includegraphics[width=0.75\linewidth]{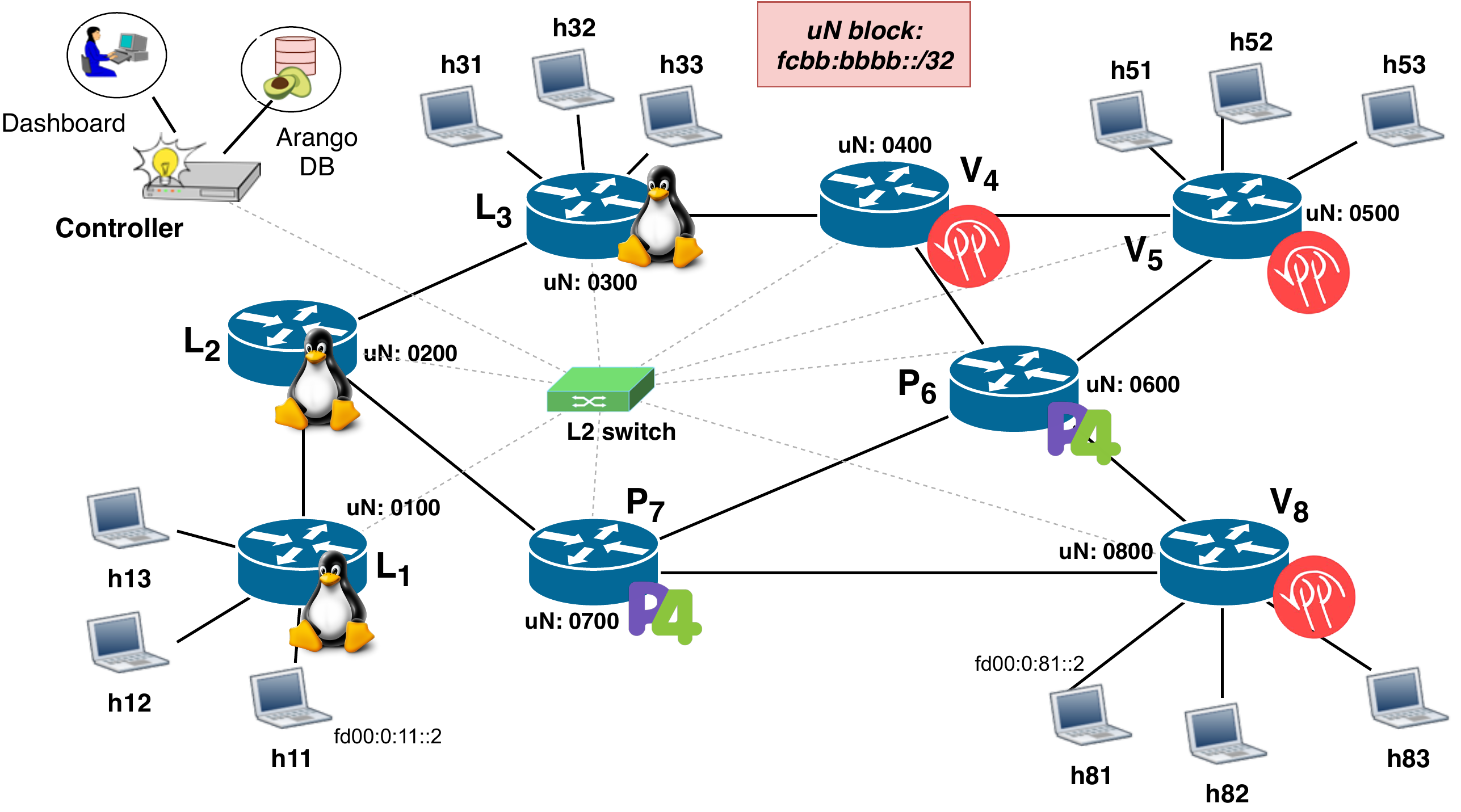}
	\caption{uN Interoperability testbed network topology}
	\label{fig:intereop_testbed}
\end{figure*}
}

\shortver{
\begin{figure}[t]
	\centering
	\includegraphics[width=\columnwidth]{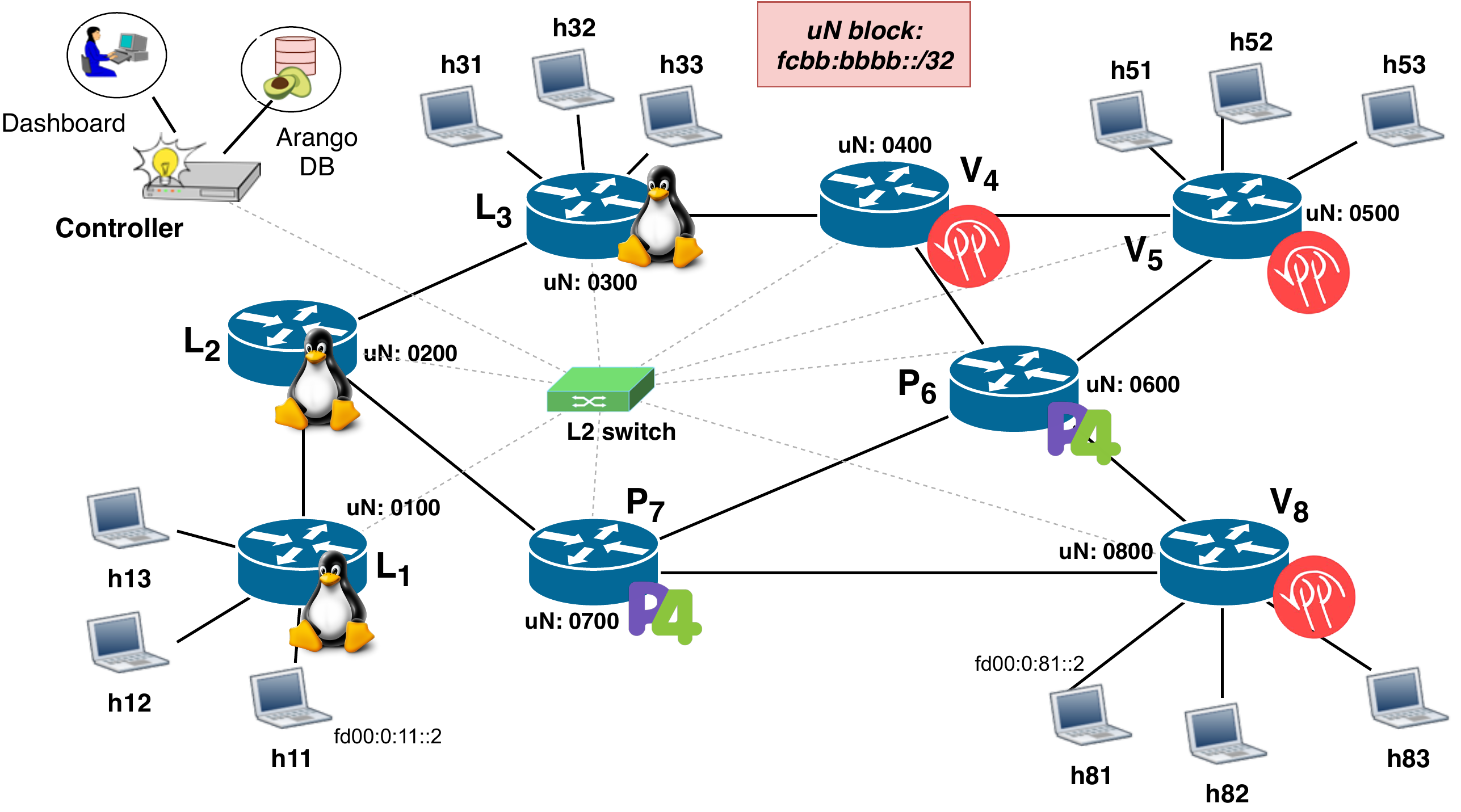}
	\caption{uN Interoperability testbed network topology}
	\label{fig:intereop_testbed}
\end{figure}
}

\ste{TODO: è giusto il simbolo rosso VPP nella figura 4? mi sembra poco comprensibile...} \tulu{E' l'unico logo (ufficiale?) di VPP che ho trovato}

\subsection{Use case description and goals}
We present a distributed use case scenario, which has two goals:
(i) provide a functional assessment of the overall header compression mechanism in a meaningful application scenario; (ii) demonstrate that the uN extension can be implemented on top of different data plane frameworks and that these different implementations are inter-operable with each others.
The demo of the proposed use case is similar to the SRv6 Micro SID Interoperability Demonstration presented by CISCO \cite{cisco_demo}. Differently from this one, our demo is reproducible and publicly available at the project repository \cite{usid-testbed-repo} and includes the detailed instructions to repeat the proposed experiments.

Figure \ref{fig:intereop_testbed} shows the network topology of the proposed use case scenario, which consists of:

\begin{itemize}
    \item 3 Linux nodes implementing the SRv6 uN functions in the kernel ($L_1$, $L_2$, $L_3$);
    \item 3 programmable data planes nodes built on top of the Vector Packet Processor platform  \cite{vpp} ($V_4$, $V_5$, $V_8$);
    \item 2 programmable data planes nodes built on top of the software-based P4$_{16}$ implementation bmv2 \cite{bmv2} ($P_6$, $P_7$); 
    \item 1 controller responsible for managing the uN dynamic configuration of paths and host traffic to be steered;
    \item 12 IPv6 enabled Linux end-hosts (h11, h12, h13, etc..).
\end{itemize}

The SRv6 uN primitive set addressed by the proposed use case scenario consists of 3 functions. The first one is the \texttt{encap} function, which is responsible for encapsulating the IPv6 legacy packet that are \dq{entering} into the SRv6 domain and specifying the uN list describing the path. This function is implemented by the edge nodes that receive packets from the transmitting end hosts. The second one is the \texttt{uN} function which is responsible for extracting the uN of the next router (as described in section \ref{sec:arch}). This function is implemented both in the intermediate nodes and in edge nodes in the SRv6 path. This function operates in two different ways, referred to as  \texttt{uN(un)} \texttt{uN(End)}. \texttt{uN(un)} consists in processing the active Micro SID and replacing it with the next one (through a shift operation). \texttt{uN(End)} consists in selecting the next SRv6 segment encoding a new micro-program, i.e. advancing the next SID in the SRH and copying it in the destination address of the IPv6 packet. This operation is performed by the uN behavior when there are no more Micro SIDs to be processed in the Micro SID container. Lastly the \texttt{decap} function is responsible for extracting the original IPv6 legacy packet sent by the transmitting end hosts. This function is implemented in the last (edge) nodes in the SRv6 path that are responsible for delivering the original IPv6 packet to the target end hosts.

\begin{table}
\centering
\resizebox{0.8\columnwidth}{!}{%
\begin{tabular}{|l|l|l|l|}
\hline
\textbf{\texttt{encap}} & \textbf{\texttt{uN(un)}}   & \textbf{\texttt{uN(End)}}       & \textbf{\texttt{decap}} \\ \hline
$L_1$, $L_3$         & $L_1$, $L_2$, $L_3$             & $L_1$, $L_2$, $L_3$             & $L_1$, $L_3$         \\ \hline
$V_5$, $V_8$ & $V_4$, $V_5$, $V_8$ & $V_4$, $V_5$, $V_8$ & $V_5$, $V_8$ \\ \hline
               & $P_7$, $P_6$           & $P_7$, $P_6$           &                \\ \hline
\end{tabular}%
}
\caption{Micro SID functions and testbed nodes}
\label{tab:uN-functions}
\end{table}

Table \ref{tab:uN-functions} summarizes the association between the SRv6 uN functions and the nodes implementing it. \ste{It is worth noting that the \texttt{encap()} and \texttt{decap()} functions can be implmented in P4, but for this demo but we chose not to implement them because ... STEFANO WHY? TODO}
 
\subsection{Testbed deployment}
As the main objective of this demo is the functional assessment of the proposed header compression mechanism (the performance assessment of the proposed implementations is realized with specific standalone experiments described in Section \ref{sec:assessment}), the use case scenario described in the previous section has been implemented in an emulated SW environment. In particular, we have designed and developed a virtual environment built on top of mininet \cite{mininet}. The relevant mininet VM includes the 3 Micro SID implementations listed in the previous section as well as the controller and the end hosts. 


\noindent \textbf{\textit{Micro SID numbering}}. 
\label{sec:usid_numbering}
For this use case we allocated the Micro SID block \texttt{fcbb:bbbb::/32}. Each node is assigned with a Micro SID in the format \texttt{fcbb:bbbb:0X00::/44}, where X is an index bound to the node in the range $[1,..,8]$. The use of the prefix length \dq{/44} instead of \dq{/48} is necessary to support the encoding of the End.DT6 directive in the least significant bit of the Micro SID (it also enables to encode 14 behaviors more). As a result, nodes with End.DT6 capability will match the \dq{/48} rule with the first bit enabled to discriminate whether to apply uN(end) or End.DT6. 

In order to solve the above mentioned issue, we implemented also an alternative solution. A special Micro SID (0xf00d) is used to support the End.DT6 and encoded at the end of the Micro SID list, e.g.  \texttt{fcbb:bbbb:0X00:f00d::}. As a result, a node supporting this feature would enable the End.DT6 action when it matches its assigned Micro SID followed by \texttt{f00d:0000::}. 

\shortver{ Further details, including the listings of the static routes configurations for Linux, P4 and VPP can be found in the extended version \cite{usid-arxiv}. 
}

\extended{
\noindent \textbf{\textit{Routing configuration}}. For sake of simplicity, the routing tables of the nodes are statically configured at startup by means of shell scripts. 
}

\extended{
The startup configuration script (which can not be listed in details for the limited space available for this publication) includes different kinds of routes:
\begin{enumerate}
  \item routes toward the ipv6 addresses of all the routers links
  \item routes toward the loopback interfaces of all the routers
  \item routes toward the end hosts IPv6 addresses
  \item routes to uN node performing the uN\_un and the uN(End) functions 
  \item routes to perform the End.DT6 decapsulation function
\end{enumerate}
}

\extended{
Examples of the above mentioned routes are provided in listings \ref{listing:linux_routes}, \ref{listing:vpp_routes} and \ref{listing:p4_routes}. It is worth noting that the 3 proposed implementations use different tools for configuring the internal routing tables: (i) the Linux kernel nodes exploit the \texttt{iproute2} suite; (ii) the VPP nodes are configured with the \texttt{vppctl} tool; (iii) the P4 nodes exploit the runtime controller client interface \texttt{simple\_switch\_CLI}. 
}

\tulu{Il gruppo di listings deve andare in extended version ma quando metto extended{} da tanti errori di compilazione (forse a causa dei listings dentro). Comunque per la short gli errori non li da.. quindi in caso scommentate la linea sotto e quella dopo i listing}
\ste{il problema del comando extended{} e' che non tollera le righe vuote... bisognerebbe ripetere extended{} ogni blocco di linee separato da una linea vuota... per ora ho messo un comment...AGGIORNAMENTO: ho risolto mettendo begin{enxtended-env}}

\begin{extended-env}

\begin{figure*}[htb]

\begin{multicols}{3}

\begin{lstlisting}[numbers=left, stepnumber=1, xleftmargin=.03\textwidth, label=listing:linux_routes,
caption={Linux routes configuration},
captionpos=b]
# Linux configuration

# link between routers
ip -6 route add fcf0:0:2:3::/64 via fe80::a41f:f2ff:fe53:d25a dev L1-L2 metric 20

# loopback interfaces
ip -6 route add fcff:2::/32 via fe80::a41f:f2ff:fe53:d25a dev L1-L2 metric 20

# end hosts
ip -6 route add fd00:0:11::/64 dev L1-h11 proto kernel metric 256

# uN behavior
ip -6 route add fcbb:bbbb:0100::/48 encap seg6local action uN dev L2-L3

# End.DT6 behavior
ip -6 route add fcbb:bbbb:0100:f00d::/80 encap seg6local action End.DT6 table 254 dev L1-h11
ip -6 route add fcbb:bbbb:0101::/64 encap seg6local action End.DT6 table 254 dev L1-h11
\end{lstlisting}

\begin{lstlisting}[numbers=left, stepnumber=1, xleftmargin=.03\textwidth, label=listing:vpp_routes,
caption={VPP routes configuration},
captionpos=b]
# VPP configuration
# link between routers
vppctl> ip route add fcf0:0:1:2::/64 via fe80::acec:4bff:fe20:4852 host-v5-p6

# loopback interfaces
vppctl> ip route add fcff:1::/32 via fe80::e078:34ff:fe41:d6ec host-v5-v4

# end hosts
vppctl> ip6 table add 100
vppctl> ip route add fd00:0:51::/64 table 100 via fd00:0:51::2 host-v5-h51

# uN behavior
vppctl> sr localsid prefix fcbb:bbbb:0500::/48 behavior un 16

# End.DT6 behavior
vppctl> sr localsid prefix fcbb:bbbb:f00d::/64 behavior end.dt6 100
vppctl> sr localsid prefix fcbb:bbbb:0501::/64 behavior end.dt6 100
\end{lstlisting}

\begin{lstlisting}[numbers=left, stepnumber=1, xleftmargin=.03\textwidth,  label=listing:p4_routes,
caption={P4 routes configuration},
captionpos=b]
# P4 configuration
# link between routers
simple_switch_CLI> table_add IngressPipeImpl.routing_v6 IngressPipeImpl.set_next_hop fcf0:0:3:4::/64 => d2:8e:4f:23:63:59

# loopback interfaces
simple_switch_CLI> table_add IngressPipeImpl.routing_v6 IngressPipeImpl.set_next_hop fcff:1::/32     => d2:8e:4f:23:63:59 

# uN behavior
simple_switch_CLI> table_add IngressPipeImpl.my_sid_table IngressPipeImpl.end_action_usid fcbb:bbbb:0700::/48 0
simple_switch_CLI> table_add IngressPipeImpl.my_sid_table IngressPipeImpl.end_action 	   fcbb:bbbb:0700::/64 0
\end{lstlisting}
\end{multicols}
\end{figure*}

\end{extended-env}

\subsection{Functional assessment: control plane operations}
In order to have a thorough interoperability assessment, we create multiple end host flows and associate each of them to different SRv6 uN enabled paths. For example, let us consider a bidirectional ICMP echo request/reply flow between the hosts \textit{h11} and \textit{h31}. For the request, the controller enforces the following path: $L_1(encap) \rightarrow L_2 \rightarrow P_7 \rightarrow P_6 \rightarrow V_5 \rightarrow V4_1(End) \rightarrow L_3(End.DT6)$. The ICMP echo reply sent by \textit{h31} matches the same path in the reverse direction. 

To express this policy from the control plane, it is just needed to trigger one command in the controller CLI that needs the following information:
\begin{itemize}
    \item the IPv6 destination address of h31, needed to install in $L_1$ the path from h11 to h31;
    \item the IPv6 destination address of h11, to install in $L_3$ the path from h31 to h11;
    \item the list of the names of nodes to traverse (in this case \texttt{l1, l2, p7, p6, v5, v4, l3}).
\end{itemize}

The controller also implements some extended features like encoding correctly the End.DT6 behavior. As an example, it supports the corner case in which the last segment of the Micro SID list contains 6 \dq{topological} Micro SIDs. In this case, there is no more space left in the destination address to insert the Micro SID expressing the End.DT6 behavior (0xf00d). It is also not allowed to create a new segment containing only the End.DT6 Micro SID (e.g. \texttt{fcbb:bbbb:f00d::}). Therefore, the controller automatically inserts 5 Micro SIDs in the first segment and in the last segment it inserts the Micro SID of the egress node followed by the End.DT6 Micro SID (e.g. \texttt{fcbb:bbbb:0300:f00d::}). It is worth noting that in the case of adopting the other type of uSID numbering described in Section \ref{sec:usid_numbering}, the entire Micro SID list would have fit inside just one IPv6 destination address, resulting in a saving of 24 bytes (8 SRH and 16 for the SID).

Other control plane features implemented for uSID include:
\begin{itemize}
    \item creating both symmetrical (same path for both outward and return packets) and asymmetrical (different paths for outward and return packets) policies;
    \item dumping the list of all installed policies;
    \item dumping a specific policy by specifying source and destination addresses of end hosts;
    \item removing a policy by specifying all the parameters or by referencing the policy ID.
\end{itemize}

\subsection{Functional assessment: data plane operations}
According to the control plane configuration described in the previous subsection, the echo request sent by h11 is intercepted by $L_1$ that performs the \texttt{encap()} function. The original ICMP packet is encapsulated in an IPv6 header with destination address \texttt{fcbb:bbbb:0200:0700:0600:0500:0400::} expressing the first half of the path. The second half of the path is encoded in the first position of the SRH SID list with address \texttt{fcbb:bbbb:0300:f00d::}. 

The encapsulated packet is then sent to $L_2$ which applies the uN\_un function, by extracting the first Micro SID (0200) and shifting the segment. The resulting SRv6 path is \texttt{fcbb:bbbb:0700:0600:0500:0400::}. The packet is then sent to the second uN node ($P_7$, identified by the Micro SID 0700). These operations are iterated until the packet reaches the last segment of the list ($V_4$) which applies the uN(End) function. Thus, $V_4$ copies the second half of the segment list in the IPv6 destination address and sends the packet to the next uN node ($L_3$). In $L_3$, acting as egress router, the packet is decapsulated and reaches the final end host h31.

For the ICMP echo reply path, the ingress node ($L_1$) encapsulates the packet encoding the uN list \texttt{fcbb:bbbb:0400:0500:0600:0700:0200::} in the IPv6 destination address and \texttt{fcbb:bbbb:0100:f00d::} in the SRv6 SID list.
The operations applied to the reply packet are analogues to the ones applied to the request and for this reason are here omitted.
\section{Performance evaluation}
\label{sec:assessment}

This section presents a performance analysis of the uN header compression mechanisms based on a set of stand alone experiments aiming at measuring the packet rate overhead introduced by the proposed extension with respect to the base SRv6 implementation. 

\subsection{Testbed deployment for the performance assessment }
\label{sec:test-deployment} 

To evaluate both the Linux kernel and the VPP uN implementation, we have reserved two bare metal servers on the federated testbed infrastructure CloudLabs \cite{cloudlab}. We have deployed a simple topology consisting of a traffic generator (TG) and a system under test (SUT). An instance of the TRex DPDK-based traffic generator \cite{trex} runs on the TG machine. \shortver{Details of the hw configuration of the two servers are in \cite{usid-arxiv}.}

\extended{The two reserved servers are x86 machines equipped with 2x Intel E5-2630 v3 85W 8C at 2.40 GHz for a total of 16 cores in hyper-threading, 32KB L1 cache, 256KB L2 cache, 20MB L3 cache, 128 GB of ECC RAM (8x 16 GB DDR4 2133 MHz PC4-17000 dual rank RDIMMs) and the Cisco network interface UCS VIC1227 VIC MLOM - Dual Port 10Gb SFP+ (connected via PCIe v3.0, 8 lanes). On the SUT machine we have installed a Linux 5.6 kernel patched with our uN behaviour implementation.}

In this simple testing scenario, we considered different bidirectional flows. Packets sent from TG are received by SUT on one network interface, processed according to the specific SRv6/Un function under measurement, and sent back to TG on the second network interface. 
We considered five experiments, each one with a specific combination of SRv6/uN function and packet type: 

\begin{enumerate}
    \item function \texttt{uN(un)} with IPv6 in IPv6 encapsulation without SRH. In this experiment the Micro SIDs are encoded directly within the destination address of the IPv6 header and the packets processed are the smallest ones of this measurement campaign (118 bytes);
    \item function \texttt{uN(un)} with IPv6 in IPv6 encapsulation without SRH. This experiment is similar to experiment 1, but the packet size is \dq{artificially} extended, by adding 40 bytes of payload padding, up to the same size of an IPv6 packet with an SRH containing two SIDs (i.e. 158 bytes);
    \item function \texttt{uN(un)} with IPv6 packets plus a SRH containing two SIDs (158 bytes);
    \item function \texttt{uN(End)} on IPv6 packets plus a SRH containing two SIDs (158 bytes);
    \item function \texttt{End} (basic SRv6) on IPv6 packets with an SRH containing two SIDs. Such behavior is considered to be our performance baseline. The other experiments are compared to this one to understand the overhead introduced by the proposed header compression mechanism. The packet size for this experiment is 158 bytes. 
\end{enumerate}{}

\subsection{Linux kernel implementation assessment}

\begin{table}
\resizebox{\columnwidth}{!}{%
\centering
\begin{tabular}{|l|l|l|l|l|}
\hline
\textbf{\#} & \textbf{Function}    & \textbf{Encap} & \textbf{PDR@0.5\%} & \textbf{Perf. Gain}\\ \hline
1             & uSID\_un & IPv6 in IPv6          &      869.61 kpps  & +2.48\%        \\ \hline
2             & uSID\_un & IPv6 in IPv6 (pad)      &      869.66 kpps   & +2.48\%              \\ \hline
3             & uSID\_un & IPv6 + SRv6          &       861.52 kpps      & +1.52\%       \\ \hline
4             & uSID\_end     & IPv6 + SRv6          &        843.17 kpps     & -0.64\%       \\ \hline
5             & End         & IPv6 + SRv6          &              848.60 kpps & \ ---------      \\ \hline
\end{tabular}
}
\caption{Linux kernel performance assessment}
\label{table:performance}
\end{table}

\ste{TODO mettere un riferimento alla metodologia di misura (paper di Ahmed su arxiv https://arxiv.org/abs/2001.06182) e al fatto che utilizziamo come metrica di prestazione il PDR@0.5\%}

The detailed results of the above described experiments for the Linux kernel uN implementation are reported in table \ref{table:performance}. For each test we performed 60 experiments with a duration of 10 seconds each. Therefore, each test is the average of the results of the 60 experiments.
The throughput (848.60 kpps) measured in the experiment 5 is taken as reference to evaluate the increase or decrease in performance experienced in the other experiments. Indeed, the SRv6 End behavior does not perform any uN operation so that it allows us to find out the impact of the uSID processing with respect to the base SRv6 processing.
For each experiment reported in table \ref{table:performance}, we run the performance tests to estimate the maximum throughput considering the Partial Drop Rate fixed at 0.5\% (PDR@0.5\%), as discussed in \cite{ahmedperformance} and \cite{abdelsalam2020srperf}.

As expected, the processing overhead introduced by the uN behavior depends on which operation is performed and on the packet encapsulation.
The IPv6-in-IPv6 encapsulation achieves the highest performance in terms of throughput. The fixed IPv6 header size along with a more efficient parsing are the key factors which increase the overall throughput of 2.48\% with respect to the baseline (SRv6 End behavior).

Considering the SRv6 encapsulation, the \texttt{uN(un)} performance is slightly better than the performance of the SRv6 End behavior with a measured gain of 1.52\%
On the other hand, when the \texttt{uN(End)} operation is applied on SRv6 packets the measured performance drop with respect to the baseline is 0.64\% and thus it could be considered practically negligible. 

These results show that the large saving in packet overhead the uSID solution provides, does not reduce performance with respect to standard SRv6 processing.

\subsection{VPP implementation assessment}
In this subsection we briefly discuss the experiment results for the VPP implementation. As expected, the packet rate measured with the VPP is one order of magnitude higher than the one obtained with the Linux kernel implementation. This is mainly due to the fact the VPP instance under measurement is built on top of DPDK\cite{dpdk}, which compared to the plain Linux kernel network subsystem performance, provides such improved overall performance. 

Indeed, for experiment 1  we measured an average packet rate of 8541.78 kpps (which,  with 118 byte packets, is close to the 10 Gbps line rate of the NIC used in the testbed). For the remaining four experiments, which are all based on 158 byte packets, we always reach the line rate, i.e. 6867.59 kpps.


\subsection{P4 implementation assessment}
As described in Section \ref{sec:design-implementation}, the implementation of uN in P4 required few lines of code and as a consequence, limited resources occupation. Moreover, taking as a reference the SRv6 P4 implementation described in \cite{srv6-cascone}, our uN solution can even reuse the table used for SRv6 processing. This brings two advantages: (i) there is no need for adding a different table for uN processing and (ii) the P4 node remains compatible with plain SRv6. In fact, from a table occupation perspective, to support uN processing it is only needed to add two entries in the table implementing the SRv6 and uN behaviors. The P4 implementation described in this paper has not been assessed in terms of performance as it is based on a behavioral model (bmv2), meant primarily for functional assessment. 

The P4 implementation presented in Section \ref{sec:design-implementation} is based on P4$_{16}$ and not compatible with existing hardware like Tofino \cite{tofino} as is. In particular the $usid\_un$ action cannot be written in P4$_{14}$ as described in the Listings reported in the extended version, but must be segmented through multiple stages.
\section{Related works}
\label{sec:related}
\subsection{SRv6 protocol extensions and optimizations}
A comprehensive survey of the research on SRv6 can be found in  \cite{ventre2019survey}. Among all the reported literature works, a considerable number is related to our work, like the ones focusing on optimizations \cite{giorgetti2015path}\cite{lazzeri2015efficient}\cite{giorgetti2015reliable}\cite{salsano2016pmsr}. A survey of the SRv6 use cases can be found in \cite{duchene-sigcomm18}.

\subsection{SRv6 header compression mechanisms}

\shortver{Several works addressing the compression of the SRv6 header have been proposed in literature. Indeed, within the IETF this problem is currently being addressed by several ongoing works  \cite{coc}\cite{bonica-spring-sr-mapped-six}\cite{bonica-6man-comp-rtg-hdr-22}\cite{decraene-spring-srv6-vlsid-03}.
}

\extended{Several works addressing the compression of the SRv6 header have been proposed in literature. Indeed, within the IETF this problem is currently being addressed by many ongoing works  \cite{li-spring-compressed-srv6-np-02}\cite{coc}\cite{bonica-spring-sr-mapped-six}\cite{bonica-6man-comp-rtg-hdr-22}\cite{decraene-spring-srv6-vlsid-03}.
}

\extended{The authors in \cite{li-spring-compressed-srv6-np-02} describe a compressed version of the SID list (C-SID) and SRH (C-SRH) that would decrease significantly the overhead carried by SRv6. Anyway, this mechanism requires some modifications in data plane, while our Micro SID solution builds upon the base SRv6 processing and needs minimal additions to existing solutions.} 

The COC solution is defined in the context of the framework called \dq{Generalized SRv6 Network Programming for SRv6 Compression} (G-SRv6) \cite{coc}. The basic idea is that in an SRv6 domain all the IPv6 SIDs can share the initial part of the address, i.e. the \textit{Locator Block} (in the uSID solution defined in the previous section, we have called it the \textit{uSID block}). Therefore it is possible to avoid carrying the full SID in the Segment List of the SRH. Only a node identifiers and a function (FUNCT) identifier is needed for each SID in the Segment List. In the COC/SRv6 solution, the first SID of the SRH is a regular SID, followed by a sequence of \dq{short} identifiers called C-SIDs (Compressed-SID). At each hop, the IPv6 Destination Address (DA) will be updated keeping the Locator Block at the beginning then inserting the C-SID (node and function identifier). The final part of the address is used to encode the pointer to the currently active C-SID identifier in the C-SID list.

Note that the two proposed solutions uSID and COC have been recently combined in the same conceptual framework in \cite{id-srh-comp-sl-enc}, wherein uSID and COC are formally defined as extensions of SRv6 End and End.X flavors. \extended{The two new flavors are respectively called NEXT-C-SID (uSID) and REPLACE-C-SID (COC) in \cite{id-srh-comp-sl-enc} and they can be used stand-alone or in their combination, referred to as NEXT-REPLACE-C-SID.}

\cite{bonica-spring-sr-mapped-six} and \cite{bonica-6man-comp-rtg-hdr-22} propose a natively compressed version of SR mapped to IPv6 (SRm6) that inserts the SID list in an extension header of IPv6, along with a 32 bit Compressed Routing Header (CRH). Although this approach provides similar compression benefits to uSID, SRm6 needs a new control and data plane, a new ecosystem (not SRv6-native) and additional lookups at egress PE \cite{dukes-srv6-overhead-analysis}.

The work described in \cite{decraene-spring-srv6-vlsid-03} proposes a mechanism to encode variable length SIDs (vSID), ranging from 1 to 128 bits, signalled by the control plane. Having SIDs of variable length increases versatility, but it comes at the cost of more complex signalling to be handled by both control and data plane.

\subsection{Segment Routing in SDN/NFV scenarios}
SRv6 has been proved to be particularly suited for SDN/NFV scenarios \cite{leb18}. Abdelsalam et al. \cite{8004208} explored the use of SRv6 for NFV service chaining. 

A widely adopted SW based implementation of SRv6 is the one provided within the Linux kernel \cite{SRV6implem}. The performance of the Linux's SRv6 implementation has been assessed in \cite{8584976}. 

Other relevant SW based implementations \cite{goldshtein2016next, Xhonneux18-ebpf} leverage the eBPF programmable data plane implemented in the Linux kernel to develop virtualized network functions. Another eBPF based SRv6 implementation has been exploited in \cite{Xhonneux18-ebpf} to realize in-network programmability use cases. Moreover, an implementation of SRv6 on P4 dataplane and ONOS controller has been presented in a tutorial\cite{srv6-cascone} and has been extended with Micro SID in \cite{europ4-demo}.
 
SR has been also exploited in SDN scenarios. \extended{Ventre et al.  \cite{8493283} presented an SDN Architecture and Southbound APIs for IPv6 Segment Routing Enabled Wide Area Network application scenarios.} Bidkar et al. \cite{bidkar2014field} presented an SDN framework built upon Carrier Ethernet and augmented with SR.
L.  Huang et al. \cite{huang2018optimizing} provide a novel SR architecture based on OpenFlow that reduces the overhead of additional flow entries and label space. Dugeon et al. \cite{dugeon2017demonstration} implement and assess the SR approach with SDN based label stack optimization on top of the SDN controller OpenDaylight.  Lee et al. \cite{lee2016efficient} propose a routing algorithm for SDN with SR that can meet the bandwidth requirements of routing requests. 
\extended{Among all works cited in this survey, a considerable number focuses on SRv6 deployment in NFV/SDN scenarios.}
\section{Conclusions}
\label{sec:conclusion}

In this paper we presented Micro SID, an extension to SRv6 that aims at reducing the protocol overhead by providing a compact representation of the segment list encoded in the IPv6 routing header (SRH).
We showed an analytic demonstration of the benefit of the proposed solution and we also proved its feasibility by providing three different open source implementations that introduce negligible processing overhead with respect to the basic SRv6 approach. In addition, we presented a reproducible interoperability demonstration of the three implementations in a meaningful distributed use case.

\section*{Acknowledgment}
This work has received funding from the Cisco University Research Program Fund and the EU H2020 5G-EVE project.

\ifCLASSOPTIONcaptionsoff
  \newpage
\fi



\bibliographystyle{IEEEtran}
\bibliography{references}

\end{document}